\documentclass[12pt,preprint]{aastex61}
\usepackage{graphicx}

\begin{document}

\shortauthors{Stuhr et al.}
\shorttitle{ARPS for Irregular Time Series}

\title{AutoRegressive Planet Search: Feasibility Study for Irregular Time Series }

\author{Andrew M. Stuhr}
\affiliation{Department of Astronomy \& Astrophysics, Pennsylvania State University, 525 Davey Lab, University Park, PA 16802}

\author{Eric D. Feigelson}
\affiliation{Department of Astronomy \& Astrophysics, Pennsylvania State University, 525 Davey Lab, University Park, PA 16802}
\affiliation{Center for Exoplanets and Habitable Worlds, Pennsylvania State University}
\affiliation{Center for Astrostatistics, Department of Statistics, 325 Thomas Building, Pennsylvania State University, University Park PA 16802}

\author{Gabriel A. Caceres}
\affiliation{Department of Astronomy \& Astrophysics, Pennsylvania State University, 525 Davey Lab, University Park, PA 16802}
\affiliation{SparkBeyond, 270 Madison Ave., Suite 702, New York NY 10016}

\author{Joel D. Hartman}
\affiliation{Department of Astrophysical Sciences, Princeton University, Princeton NJ 08544}

\correspondingauthor{Eric D. Feigelson}
\email{edf@astro.psu.edu}

\begin{abstract}
Sensitive signal processing methods are needed to detect transiting planets from ground-based photometric surveys. Caceres et al. (2019) show that the AutoRegressive Planet Search (ARPS) method --- a combination of autoregressive integrated moving average (ARIMA) parametric modeling, a new Transit Comb Filter (TCF) periodogram, and machine learning classification --- is effective when applied to evenly spaced light curves from space-based missions. We investigate here whether ARIMA and TCF will be effective for ground-based survey light curves that are often sparsely sampled with high noise levels from atmospheric and instrumental conditions. The ARPS procedure is applied to selected light curves with strong planetary signals from the Kepler mission that have been altered to simulate the conditions of ground-based exoplanet surveys. Typical irregular cadence patterns are used from the HATSouth survey. We also evaluate recovery of known planets from HATSouth. Simulations test transit signal recovery as a function of cadence pattern and duration, stellar magnitude, planet orbital period and transit depth.   Detection rates improve for shorter periods and deeper transits. The study predicts that the ARPS methodology will detect planets with $\gtrsim 0.1$\% transit depth and periods $\lesssim 40$ days in HATSouth stars brighter than $\sim$15 mag. ARPS methodology is therefore promising for planet discovery from ground-based exoplanet surveys with sufficiently dense cadence patterns.  
\end{abstract}


\section{Introduction \label{intro.sec}}

A major impediment to detecting transiting exoplanets in stellar photometric light curves is the presence of autocorrelated nuisance signals either from the star itself or from the observing conditions \citep{Tamuz05, Pont06, Southworth08}. Space-based surveys are mostly affected by magnetic activity or other sources of stellar variability that can hide planetary transits, while ground-based surveys suffer contaminating noise components from instrumental effects and changes in atmospheric conditions.   Most past efforts to remove these unwanted signals use nonparametric statistical models such as moving averages \citep{Zhang18}, Gaussian Processes or other local regression procedures \citep{Gibson12, Haywood14, Aigrain16, Luger16}, wavelet analysis \citep{Carter09, Jenkins10}, and advanced signal processing methods like Independent Component Analysis \citep{Waldmann12}, singular spectrum analysis \citep{Boufleur18}, and correntropy \citep{Huijse12}.   

It is also reasonable to try low-dimensional parametric models based on stochastic autoregressive processes.  For stationary time series where statistical properties such as mean and variance are unchanging throughout the observations, a common procedure fits autoregressive moving average (ARMA) models that are effective at treating a very wide range of aperiodic short-memory autocorrelated behaviors.  Since many astronomical data are non-stationary, a differencing operator is needed to remove trends;  these fits are called ARIMA models. When the differencing operator is allowed to be fractional, it becomes an ARFIMA model that parametrizes a long-memory $1/f^\alpha$ `red noise' process as well as short-memory autocorrelation. These fits are obtained by maximum likelihood estimation ARIMA-type modeling with model selection based on the Akaike Information Criterion, where likelihood improvements are balanced with model parsimony. ARIMA-type model is presented in detail in textbooks such as \citet{Chatfield03, Box15, Hyndman15}; applications to time domain astronomy are discussed by Feigelson et al. (2018). 

Caceres et al. (2019a,b, henceforth Papers I and II) show that ARIMA and ARFIMA work well for modeling the variability of light curves from NASA's 4-year Kepler mission while leaving periodic planetary transit signals mostly intact. The differencing operation changes a box-shaped transit into a double-spike, so a new Transit Comb Filter (TCF) periodogram is developed to replace the commonly used Box-Least Squares algorithm \citep[BLS,][]{Kovacs02}.  The combination of ARIMA, TCF, and (when training sets are available) machine learning classification to reduce False Positive from non-planetary periodic signals is nicknamed the {\it AutoRegressive Planet Search} or ARPS procedure.  ARMA modeling is designed for evenly spaced data, though missing data points are often permitted. The Kepler missing produced photometric measurements every 29.4 minutes for $\sim$4 years with $\sim$15-20\% missing data due to satellite operations.  

However, applying autoregressive modeling to ground-based data is less obvious.  Ground based data from a single telescope will have $\sim$16 hour gaps every day due to daylight as well as $\sim$6 month gaps every year.  Cloudy weather, instrumental problems, and telescope scheduling allocations can create additional gaps. The resulting observational cadence patterns can be extremely irregular with greater than 90\% missing data when placed onto an evenly-spaced time grid.  Instrumental and atmospheric variations can dominate stellar and planetary signals, and it is not clear that these effects will be well-modeled by low-dimensional ARIMA-type models. 

Irregular time series arise in a number of fields but the great bulk of time series analysis is designed for regularly spaced data.   General discussion of methodology for irregular spacing appears in \citet{Parzen83}.  A common treatment is to interpolate the data onto an evenly spaced grid and then use standard methods \citep{Gentili04} and to `impute' values for missing time \citep{Moritz17}.  Many traditional methods accept evenly spaced time series with `missing data': ARIMA modeling, Kalman filtering, Ljung-Box test for autocorrelation, and so forth.  For some parametric models, such as fitting sinusoidal functions for periodic behaviors with the Lomb-Scargle periodogram or autoregressive functions for aperiodic behaviors with continuous-time ARMA-type models, specialized procedures have been developed that maintain the irregular cadences \citep[e.g.,][]{Scargle82, Brockwell01}.  

This question is addressed in the present study.  We investigate the effectiveness of the multistage ARPS procedure developed in Paper I for  transiting planet detection using ground-based photometric transit surveys characterized by irregular, sparse cadences and noise components from stellar variability and atmospheric effects.   Observational cadence patterns associated with the HATSouth project of the 3-telescope HATNet (Hungarian Automated Telescope Network) are chosen to illustrate the findings.  Our analysis also provides guidance regarding optimal observational strategies for future transit surveys. 

For this study, we test these statistical procedures using stars with known existing planets, either from the Kepler mission or from the HATSouth survey, in conjunction with cadence patterns and noise characteristics from ground-based surveys.  Our procedure is to take Kepler light curves for stars with strong planetary signals, prune them to a typical ground-based cadence pattern, redistribute the data onto an evenly spaced grid, add simulated ground-based noise, fit ARMA-type models, and search for periodic transits in the model residuals using TCF periodogram developed in Paper I.   We examine the sensitivity of ARPS processing on various parameters such as survey duration, cadence pattern, star brightness, planet transit depth, and period.  

The result of our study is that the ARPS method has reasonable sensitivity compared to past analysis procedures for transiting planet detection using BLS.   Since the mathematical foundations of parametric ARIMA-type and previous nonparametric procedures for treating unwanted sources of variability are very different, each method may capture different true planetary signals.  It is quite possible that ARPS is less sensitive for planet detection than established methods for some light curves, but more sensitive for other light curves.   Its application to ground-based surveys with sufficiently dense candences thus seems warranted.

\section{Statistical Methodology}

\subsection{Overview of the AutoRegressive Planet Search Methodology \label{Method_overview.sec}}

ARPS methodology requires an evenly spaced light curve; treatment of unevenly spaced data is covered in \S\ref{Method_irreg.sec}. First, the light curve data are fitted to a low-dimensional ARIMA-type model.  In autoregressive modeling, current values of a time series are functionally dependent on past values; this differs from other regression procedures where current values are functionally dependent on time.  The ARIMA model, promoted by Box and Jenkins in the 1970s, has proven to be highly flexible and effective for many forms of stochastic variability.  It has three types of linear dependencies: AR for autoregressive, I for integrated, and MA for moving average.  We review the mathematics here; more details are provided in Paper I with authoritative treatments in texts like \citet{Hamilton94}, \citet{Chatfield03}, and \citet{Box15}. 

The AR component is formulated
\begin{equation} \label{AR}
x_t = \phi_1 x_{t-1} + \phi_2 x_{t-2} + ... + \phi_p x_{t-p} + \epsilon_t
\end{equation}
where $x_t$ is the value at a selected time bin $t$, $p$ is the order of the model, $\epsilon_t$ is a Gaussian error component, and $\phi$ is a vector of lag coefficients with length $p$. The parameters $\phi_i$ can be calculated using least squares or maximum likelihood estimation. The order $p$ is selected using a penalized likelihood measure such as  the Akaike Information Criterion. 

While the AR component focuses on how recent previous data values affect future ones, the moving average component describes how data points are affected by previous random disturbances. This dependence is formulated 
\begin{equation} \label{MA}
x_t = \epsilon_t + \theta_1 \epsilon_{t-1} + \theta_2 \epsilon_{t-2} + ... + \theta_q \epsilon_{t-q}
\end{equation}
where $q$ is the order, $\epsilon_t$ is the Gaussian error at time $t$, and $\theta$ is the vector of lag coefficients with length $q$.  Equations \ref{AR}-\ref{MA} are combine to create an ARMA(p,q) model represented by 
\begin{equation} \label{ARMA}
x_t = \epsilon_t + \sum_{i=1}^{p} \phi_i x_{t-i} +  \sum_{j=1}^{q} \theta_j \epsilon_{t-j}.
\end{equation}

However, ARMA models require stationary behaviors whereas many time series in astronomy, particularly stellar light curves, are non-stationary due to trends in the mean value. The differencing operator treats many forms of non-stationarity, represented by
\begin{equation} \label{backshift}
(1-B)^d x_t = \epsilon_t  ~~ {\rm where}~~ Bx_t = x_{t-1}
\end{equation}
where $d$ is the order of differencing and $B$ is the backshift operator. Often, $d=1$ is sufficient for converting a non-stationary time series into a stationary one. Here, the time series $x_t$ is replaced by the time series $x_t - x_{t-1}$.  For integer values of $d$, the notation for this operation is 'I' because the original time series can be recovered by integrating the new time series.  

When $d$ is a fraction rather than an integer, the model is called ARFIMA: autoregressive fractionally integrated moving average.  This generalization give a parametric form for long-memory processes that is equivalent to $1/f$-type `red' noise that is well-known in astronomy.  The fitted value of $d$ is arithmetically related to $\alpha$ for $1/f^\alpha$ noise, $\alpha = 2d$.  The mathematics of this nonlinear model is complicated \citep{Palma07}. 

Summing the differencing operator (equation \ref{backshift}) with the ARMA model (equation \ref{ARMA}) gives the following ARIMA(p,d,q) or ARFIMA(p,d,q) model:
\begin{equation} \label{ARIMA}
(1-B)^d x_t = \epsilon_t + \sum_{i=1}^{p} \phi_i x_{t-i} +  \sum_{j=1}^{q} \theta_j \epsilon_{t-j} 
\end{equation} 

Once a stellar light curve is fit with an ARIMA, ARFIMA, or similar model, the model is subtracted from the original data and the periodic transit signal is sought in the residuals.  Because a box-like shape is converted into a pair of spikes by the differencing operation, Paper I developed the TCF matched filter algorithm to measure the correlation of a periodic sequence of spikes in the autoregressive fit residuals: first a downward facing spike corresponding to the planetary ingress, then an upward facing spike corresponding to the egress. The spike intensity is plotted for a predetermined list of periods, phases, and durations to generate a periodogram.  If the periodogram shows one or more strong peaks, then periodic transit-like signals may be present in addition to the autoregressive process. As the TCF ignores the middle of the transit, it is less sensitive than BLS for transits with long durations and (in many cases) long periods.  But, we find in Paper II that the TCF is often comparably sensitive or more sensitive than a BLS-based analysis for shorter periods.

As with all frequency domain periodicity searches, it is difficult to evaluate the statistical significance of TCF peaks; periodogram noise is very non-Gaussian and simple 3-sigma-type criteria are ineffective.   After removing large-scale trends in the TCF periodogram, we create a measure of signal-to-noise ratio (SNR) by dividing the peak power by the median absolute deviation (MAD) obtained within the local neighborhood of a period of interest.  These steps are presented in Paper I.  

In the present feasibility study, we do not treat the third and fourth stages of ARPS processing involving a ARIMAX (ARIMA with exogeneous variables) fit and application of a Random Forests classifier trained to samples of known planets.  In this last stage, several dozen scalar `features' would be collected from different stages of analysis and trained positively to a sample of astronomical confirmed (or at least probable) planet candidates, and trained negatively to samples of random stars and known contaminants such as eclipsing binaries. ROC curves plotting True Positive and False Positive recovery rates would be examined, a threshold in Random Forest probability would be assigned, and new candidate stars satisfying this criterion would be vetted.

\subsection{Treatment of Irregularly Spaced Light Curves \label{Method_irreg.sec}}

ARIMA-type models require evenly spaced time series but often permit the presence of `missing data' for some time slots.  We apply a simple algorithm to convert an irregularly spaced time series into a regularly spaced time series with missing data.  First, all time values are shifted such that the first observation starts at time zero. The irregularly spaced data are then binned into a chosen bin width.   In our experimentation, reasonable choices of bin width had little effect on the results; periodogram signal was reduced only when the bin width is much shorter than the ingress time.  The results below are based on 29.4 min bins characteristic of the Kepler dataset.  When multiple observations fall into one bin, the values are averaged. Empty bins are treated as missing data, registered as `NA' or `Not Available' in R syntax. In this process, uncertainties of individual measurements are ignored and the resulting time series becomes heteroscedastic due to the variation in contributing data points for each bin. This effect is not considered in this study; instead each binned data point is given equal weight in the ARIMA modeling. Once the data are binned, the resulting data set is an evenly spaced time series with missing values, ready for ARIMA-type modeling.  

For methods that do not permit missing data, we replace the NA values by zero, noting that the differencing operation (equation 4) brings the time series to zero mean.  This is a simple and commonly used method for dealing with missing data gaps of varying length in zero-mean time series \citep{Gentili04}.  

More elaborate imputation of missing data than simple averaging within bins has been applied to convert irregular light curves into regularly spaced light curves.  The ARIMA model itself could be applied as in studies by  \citet{Fahlman82}, \citet{Pascual15} and \citet{ Moritz17}.  \citet{Hanif15} interpolate the light curve with cubic Hermite polynomials and then apply ARIMA models to a regularly spaced sample of the continuous estimator.  

\subsection{Simulating Ground Based Noise \label{Method_mag.sec}}

When measured with high precision ground-based instrumentation, star brightness as measured by ground-based instrumentation is subject to various sources of noise, some of which have clear time-dependent autocorrelation.  These noise contributions can severely impact the ability to extract a planetary transit signal.  Based on analysis of large ensembles of HATSouth stellar observations from telescopes in Chile, Namibia and Australia, we model the noise in two components: Gaussian white noise and autoregressive noise.  These arise from a combination of atmospheric and instrumental effects.  

{Two corrections to HATSouth magnitudes are applied in advance of ARIMA modeling \citep{Kovacs05, Bakos10}. External Parameter Decorrelation (EPD) treats outliers and simple dependencies on sub-pixel position, detector background, stellar point spread function, hour angle and zenith distance. The Trend Filtering Algorithm (TFA) removes variations shared by an ensemble of nearby stars, presumably arising from changing atmospheric conditions. The result of EPD and TFA corrections is that the majority of HATSouth stars exhibit much reduced autocorrelation compared to the raw light curve. However, the noise level within a given light curve can vary considerably due, for example, to the presence of a bright moon. In statistical parlance, the typical HATSouth light curve after EPD and TFA corrections is heteroscedastic but with reduced autocorrelation.

The empirical relationship of variance of stellar magnitude in HATSouth data after removal of collective effects with EPD and TFA can be fit with a nonlinear function
\begin{equation} 
\label{noise}
\sigma^2 = 10^{0.4(m-24)} + 10^{4+0.8(m-24)} + 0.003^2 * e^{-\frac{|t_i - t_j|}{0.02}}
\end{equation} 
in units of mag$^2$.  The first term gives shot noise from the star, and the second term represents noise from the background sky. In these terms, $m$ is the average visual magnitude determined by the relation $log(f) = 10.14 - 0.4m$ for some Kepler Pre-Data Conditioning flux $f$.  This noise component is assumed to be Gaussian with zero mean. The third term is an assumed autoregressive component arising from uncorrected atmospheric conditions.  The functional form is derived from examination of HATSouth time series: an exponentially decaying variance with characteristic timescale of 0.02 day (0.5 hr) and amplitude of 3 mmag.\footnote{As both space-based and ground-based surveys are discussed in this study, depth is various quantified as a flux fraction (ppm = parts per million) and as a magnitude difference (mmag = milli-magnitude). The two units scale approximately linearly where 10,000 ppm is equivalent to 10 mmag.}   Here the times  $t_i$ and $t_j$ are measured in days.

The Gaussian noise components and the autocorrelated noise component in equation (\ref{noise}) are added to the magnitude or flux of a stellar light curve to simulate the ground-based noise components. For example, an 8th magnitude star would have simulated Gaussian noise with a standard deviation of 0.003 magnitudes while a 14th magnitude star would have Gaussian noise with a standard deviation of 0.014 magnitudes. The autoregressive noise adds an additional 0.003 magnitudes to all stars with significant autocorrelation up to $\sim 1$ hour.

We note that each ground-based survey will have its own characteristic noise properties based on instrumental and environmental conditions. The results here are applicable to HATSouth and should be viewed as approximate for other surveys. For example, the atmospheric conditions at Dome C in Antarctica may be better than at HATSouth telescope locations, so the noise characteristics may be improved \citep{Zhang18}.

\subsection{Implementation of R for ARPS Methodology \label{R_Imp.}}

We implement the methodology described in \S\ref{Method_overview.sec} and Paper I using the public domain R statistical software environment \citep{R17}. Details are given in Paper II, and briefly summarized here.  After binning onto a regular time grid (\S\ref{Method_irreg.sec}), the noise components are added to the light curve using R's {\it rnorm} function for Gaussian noise and R's {\it arima.sim} function for the autocorrelated noise with an AR(3) model $\phi = [23, 0.9, 0.1]$.  This model, obtained with CRAN package {\it  FitAR}, mimics the behavior of equation 7 that matches typical HATSouth noise behavior.  Then the {\it diff} function produces a differenced light curve to ensure that the order is at least $d=1$ to match the shape required by the TCF algorithm. The {\it auto.arima} function in the {\it forecast} package \citep{forecast17} fits an ARIMA model with automatic model selection based on the Akaike Information Criterion. Model order is restricted to be smaller than ARIMA(5,1,5), and the computationally efficient {\it stepwise} iteration procedure is used for model selection.

The ARIMA model residuals obtained with the {\it residuals} function are input into the TCF Fortran function to calculate the TCF periodogram. The periods examined range from 0.2 days to 500 days with density coverage inversely proportional to period (see Paper I). The output of the TCF function contains values for TCF strength with associated durations and phases for a preselected list of trial periods. The ARIMA fitting and TCF periodogram construction typically take a few CPU minutes for light curves with $\sim 10,000$ brightness measurements.  

\section{Datasets and Cadences for ARPS Testing}

Two datasets are used for the analysis here: stars with unusually deep transits in NASA's 4-year Kepler mission \citep{Borucki10}, and stars with confirmed transits from the HATnet HATSouth survey \citep[HATS,][]{Bakos13}). The Kepler data are long-cadence light curves from Data Release 25 for Quarters 1 through 17 obtained from the Kepler Data Products residing at NASA's Mikulski Archive for Space Telescopes (MAST) with the final catalog provided by \citet{Thompson18}. 

Table \ref{Kepler19.tbl} with 19 entries shows all stars from the set of confirmed Kepler planet hosts with unusually strong transits satisfying the following criteria: transit depths $>$5000 ppm, orbital periods $1 < P < 50$ days, and stellar brightnesses $10 < Kepmag < 15$ mag. The depths are expressed in units of mmag for compatibility with the HATSouth stars. They are listed in order of increasing period. 

The Kepler light curves typically have 4 years duration with an evenly spaced 29.4 minute cadence and 15$-$20\% missing data due to satellite operations. Fluxes after after Pre-Data Conditioning (PDC flux) analysis are used. Planet periods and depths were obtained from NASA's Mikulski Archive for Space Telescopes2 (MAST) in summer 2017. The final column of Table \ref{Kepler19.tbl}  is a result from our simulations of ground-based surveys (\S\ref{Detectability.sec}).

Table \ref{HATS34.tbl} lists thirty-four stars with confirmed planets provided by the HATSouth team\footnote{\url{https://hatsouth.org/planets/}}.   The HATSouth survey (Bakos et al. 2013) uses six telescope units to conduct wide-field photometric time-series observations of the sky in search of transiting exoplanets. The telescopes are located at three observatories in the southern hemisphere (Las Campanas Observatory in Chile, LCO; the HESS gamma ray telescope site in Namibia, HESS; and Siding Spring Observatory in Australia, SSO) with two telescope units at each site. Each unit produces a $8\fdg 2 \times 8\fdg 2$ mosaic image at a scale of $3\farcs 7$\,pixel$^{-1}$.  

At any given time each telescope unit is assigned to observe one of 838 discrete pointing positions (fields) used to tile the celestial sphere. This primary field will be observed by the unit continuously throughout the night. Each telescope unit is also assigned a secondary field to be monitored when the primary field is below 30$^{\circ}$ altitude, or when the moon is located within the primary field. The angular distance between neighboring fields is smaller than the image field of view, so that some sources may be observed in multiple fields. Typically two primary fields are monitored at any given time by the network, with one field assigned to a set of telescope units at each of the three observatories.  This enables round-the-clock monitoring of the sky during the southern winter, so long as the weather conditions permit observations at each of the sites. Observations are collected continuously between evening and morning $12^{\circ}$ twilight at a cadence of 4 minutes, so long as a number of weather conditions are met (i.e., no precipitation, the wind and humidity are below set thresholds, and the extinction from clouds is below a set threshold). Occasional data gaps may also be present due to instrument failures or servicing. 

Through 2017 the LCO, HESS and SSO sites have averaged 8\,hrs, 7.5\,hrs and 5.5\,hrs of useful dark time per 24\,hr time period, respectively \citep{Bakos18}.  Typically a field is observed by the network for a two to six month period when it is most visible at night. When a field culminates at midnight it can be observed for much (in some cases all) of the night from a single site, while at the start and end of the season the field may only be observed for a few hours each night at a given site. Some fields have been revisited in subsequent years, with the total amount of time elapsed between revisits varying from one to as many as six years. The total number of observations gathered for a given field ranges from 2,000 to more than 37,000 images with a median value of 10,000. The median total time span for a given field is 8 months. The total duty cycle for the fields varies from 0.3\% (a secondary field with $\sim 2000$ observations gathered over 6 years), to 38\% (a field with 37,000 images and a time span of 9 months) with a median value of 10\%. 

\begin{deluxetable}{lllrrrr}
\tablecaption{Kepler Stars with Deep Transits\label{Kepler19.tbl}}
\tabletypesize{\footnotesize}
\tablewidth{0pt}
\centering
\tablehead{
\colhead{KIC} & \colhead{Name} & \colhead{Discovery} & \colhead{Period} & \colhead{Depth} & \colhead{KepMag} & \colhead{Threshold} \\
 & & & \colhead{day} & \colhead{mmag} & \colhead{mag} & \colhead{mag}
}
\startdata
010619192 & Kepler-17b & \citet{Desert11} & 1.486 & 20.8~~ & 14.10~~ & $>16$~~~~ \\
004570949 & Kepler-76b & \citet{Faigler13} & 1.545 & 5.82 & 13.30~~ & 14.0~~ \\
003749365 & Kepler-785b & \citet{Morton16} & 1.974 & 30.2~~ & 15.71~~ & $>16$~~~~\\
010666592 & Kepler-2b & \citet{Southworth11} & 2.205 & 6.67 & 10.46~~ & 13.3~~ \\
005794240 & Kepler-45b & \citet{Southworth11} &2.455 & 36.9~~ & 16.88~~ & $>16$~~~~ \\
011517719 & Kepler-840b & \citet{Morton16} & 2.496 & 25.1~~ & 14.15~~ & $>16$~~~~ \\
009651668 & Kepler-423b & \citet{Endl14} & 2.684 & 18.1~~ & 14.29~~ & $>16$~~~~ \\
010874614 & Kepler-6b & \citet{Dunham10} & 3.235 & 10.6~~ & 13.30~~ & 15.8~~ \\
010019708 & Kepler-490b & \citet{Morton16} & 3.269 & 10.1~~ & 14.88~~ & 15.7~~ \\
005358624 & Kepler-428b & \citet{Hebrard14} & 3.526 & 22.3~~ & 15.40~~ & $>16$~~~~ \\
011804465 & Kepler-12b & \citet{Fortney11} & 4.438 & 16.4~~ & 13.80~~ & $>16$~~~~ \\
011187436 & Kepler-957b & \citet{Morton16} & 5.907 & 5.57 & 15.66~~ & \nodata~~ \\
000757450 & Kepler-75b & \citet{Hebrard13} & 8.885 & 16.1~~ & 15.26~~ & 15.1~~ \\
005972334 & Kepler-487b & \citet{Morton16} & 15.359 & 14.6~~ & 14.99~~ & 15.5~~ \\
002987027 & Kepler-489b & \citet{Morton16} & 17.276 & 10.5~~ & 14.02~~ & 15.3~~ \\
003323887 & Kepler-9b & \citet{Holman10} & 19.271 & 6.66 & 13.80~~ & \nodata~~ \\
006061119 & Kepler-699b & \citet{Morton16} & 27.808 & 25.4~~ & 15.48~~ & \nodata~~ \\
011449844 & Kepler-468b & \citet{Morton16} & 38.479 & 23.5~~ & 13.78~~ & 15.7~~ \\
006522242 & Kepler-706b & \citet{Morton16} & 41.408 & 22.8~~ & 15.20~~ & 15.6~~ \\
\enddata
\end{deluxetable}

\begin{deluxetable}{lllllclllll}
\tablecaption{HATSouth Stars with Confirmed Planets \label{HATS34.tbl}}
\tabletypesize{\tiny}
\tablewidth{0pt}
\centering
\tablehead{
\colhead{Name} & \colhead{Discovery} & \colhead{Period} & \colhead{Depth} & \colhead{Mag} &\colhead{~~}& 
\colhead{Name} & \colhead{Discovery} & \colhead{Period} & \colhead{Depth} & \colhead{Mag}\\
 & & \colhead{days} &  \colhead{mmag} & && & & \colhead{days} &  \colhead{mmag} &
}
\startdata
HATS-2 b & \citet{Mohler13} & 1.3541 & 19.3 & 13.3 && HATS-19 b & \nodata & 4.56967 & 10.3 & 12.8 \\
HATS-3 b & \citet{Bayliss13} & 3.57485 & 11.1 & 11.9 && HATS-20 b & \nodata & 3.79930 & ~8.8 & 13.6 \\
HATS-4 b & \citet{Jordan14} & 2.51673 & 13.9 & 13.2 && HATS-21 b & \nodata & 3.55440 & 13.9 & 12.1 \\
HATS-5 b & \citet{Zhou14} & 4.76339 & 12.6 & 12.5 && HATS-22 b & \citet{Bento17} & 4.72281 & 22.1 & 13.1 \\
HATS-6 b & \citet{Hartman15} & 3.32527 & 35.1 & 13.9 && HATS-23 b & \citet{Bento17} & 2.16052 & 27.4 & 13.7 \\
HATS-7 b & \citet{Bakos15} & 3.18532 & ~5.5 & 12.9 && HATS-24 b & \citet{Bento17} & 1.34849 & 18.5 & 12.6 \\
HATS-8 b & \citet{Bayliss13} & 3.58389 & ~7.2 & 14.2 && HATS-25 b & \citet{Espinoza16} & 4.29864 & 14.8 & 12.9 \\
HATS-9 b & \citet{Brahm15} & 1.91531 & ~5.7 & 13.0 && HATS-26 b & \citet{Espinoza16} & 3.30238 & ~8.4 & 13.0 \\
HATS-10 b & \citet{Brahm15} & 3.31285 & ~8.9 & 12.9 && HATS-27 b & \citet{Espinoza16} & 4.63704 & ~8.7 & 12.7 \\
HATS-11 b & \citet{Rabus16} & 3.61916 & 12.6 & 13.8 && HATS-28 b & \citet{Espinoza16} & 3.18108 & 19.2 & 13.7 \\
HATS-12 b & \citet{Rabus16} & 3.14283 & ~4.3 & 12.7 && HATS-29 b & \citet{Espinoza16} & 4.60587 & 15.6& 12.5 \\
HATS-13 b & \citet{Mancini15} & 3.04405 & 21.3 & 13.7 && HATS-30 b & \citet{Espinoza16} & 3.17435 & 14.0 & 12.2 \\
HATS-14 b & \citet{Mancini15} & 2.76676 & 14.2 & 13.8 && HATS-31 b & \citet{deValBorro16} & 2.54955 & ~9.0 & 12.9 \\
HATS-15 b & \citet{Ciceri16} & 1.74749 & 16.4 & 14.6 && HATS-32 b & \citet{deValBorro16} & 2.81265 & 14.9 & 14.2 \\
HATS-16 b & \citet{Ciceri16} & 2.68650 & 12.5 & 13.7 && HATS-33 b & \citet{deValBorro16} & 2.54955 & 16.6 & 11.8 \\
HATS-17 b & \citet{Brahm16} & 16.25461 & ~5.7 & 12.1 &&  HATS-34 b & \citet{deValBorro16} & 2.10616 & 24.4 & 13.7 \\
HATS-18 b & \citet{Penev16} & 0.83784 & 19.7 & 13.9 && HATS-35 b & \citet{deValBorro16} & 1.82099 & 12.0 & 12.4 \\
\enddata
\end{deluxetable}

\begin{figure}
\centering
\includegraphics[width=1.0\textwidth]{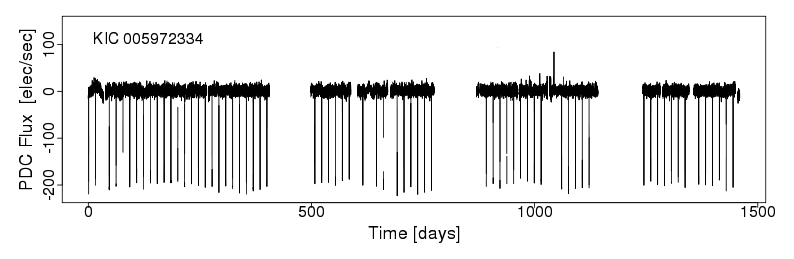}
\includegraphics[width=1.0\textwidth]{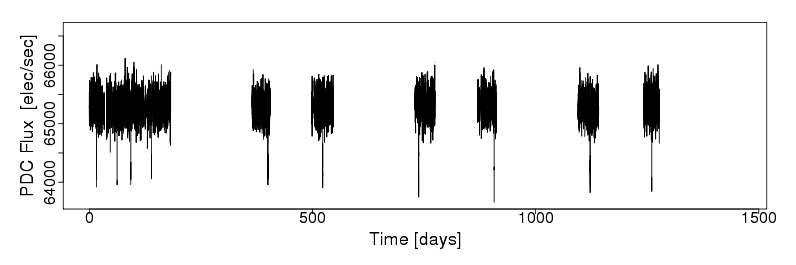}
\includegraphics[width=1.0\textwidth]{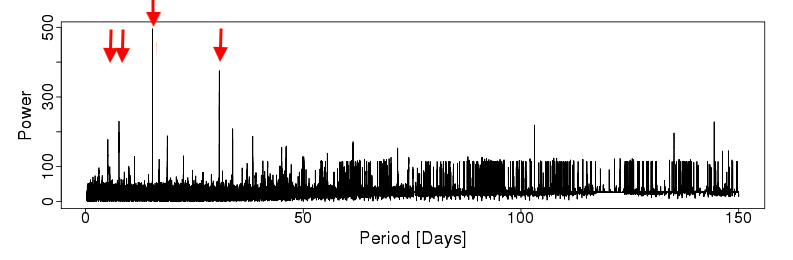}
\caption{Kepler planet with ground-based cadence pattern.  The top panel shows the original light curve for KIC 005972334 showing a large transiting planet with period $P=15.4$ days. The robust InterQuartile Range noise level is $IQR=7$. The middle panel shows the light curve simulated at 12th magnitude with added ground-based noise, 6 month continuous gaps removed, and cadence pattern reduced to 8 hours of observation per day.  Here $IQR=300$.  The bottom panel shows the TCF periodogram for the simulated light curve in the middle panel with the known period recovered along with harmonics marked with red arrows.   \label{example_sim.fig}}
\end{figure}

\section{Detectability of Kepler Planets in Ground-based Surveys}
\section{Discoverability of Kepler Planets with Ground-based Cadences} \label{Detectability.sec}

Our first method of testing ARPS methodology modifies Kepler data to mimic data taken by a ground-based survey. We first remove six continuous months of data from each year to simulate conditions on the ground for when the star is not visible.  Parameters such as the apparent magnitude of the star and cadence --- 8 hours, 16 hours, or 24 hours per day --- are then adjusted. The resulting cadence densities range from an NA fraction of 50\% to 83\%. A planet is viewed as `recovered' when the TCF power signal-to-noise ratio (SNR) of the correct period is $SNR > 20$ with respect to a window of 25,000 periods on each side of the chosen period.  Recovered periods must match published periods to better than 0.01\% accuracy. Figure \ref{example_sim.fig} shows recovery of a transit signal for for a Kepler star with period $\simeq 15$ day and depth $\simeq 0.15$\% simulated at 12th magnitude observed 8 hours per day for four seasons. 

\begin{figure}
\centering
\includegraphics[width=1\textwidth]{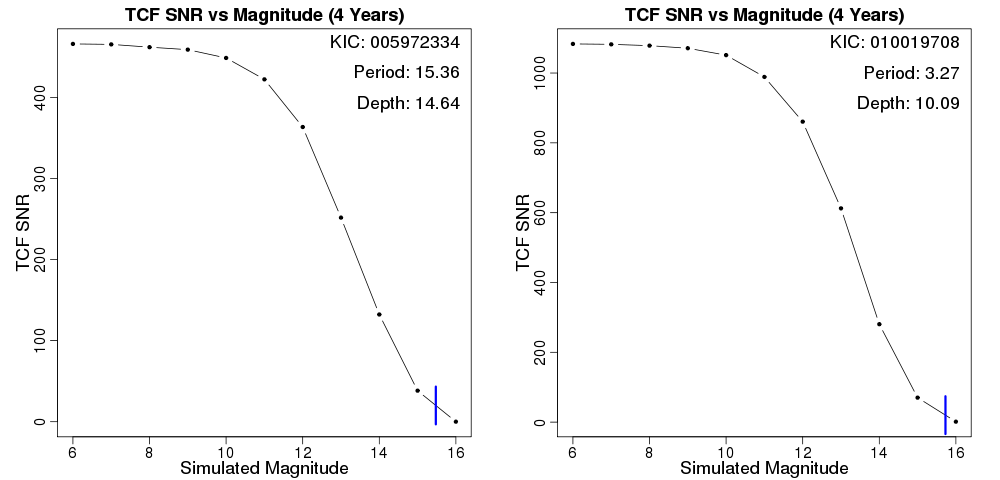}
\caption{TCF peak signal-to-noise ratio dependence on simulated magnitude for two Kepler stars with transit depth in mmag and period in days: (left) KIC 005972334 with longer period and stronger depth; (right) KIC 010019708 with shorter period and shallower depth. The blue vertical lines show the magnitude where SNR drops below 20.  \label{Mag.fig}}
\end{figure}

\subsection{Magnitude Dependence}\label{Detect_man.sec}

The noise parameters introduced in \S\ref{Method_mag.sec} indicate that noise from ground-based factors will be small for brighter stars but will dominate for fainter stars, diluting the transit signal in the TCF periodogram. For the simulated ground-based observations of the Kepler systems listed in Table \ref{Kepler19.tbl}, we varied the apparent magnitudes from $m = 6$ to $m = 16$ to quantify this signal degradation. For each test, SNR values of signal are recorded for resulting TCF periodogram peaks at periods corresponding to the known orbital period for a planet around the Kepler star. The light curve duration is kept at 4 years for this test of dependence on stellar brightness.

The simulation results for KIC 005972334 $-$ the long period, deep transit star shown in Figure~\ref{example_sim.fig} $-$ are shown in the left panel of Figure \ref{Mag.fig}.  The TCF power SNR (and therefore planet recovery) strongly decays as magnitude increases. Brighter than 9th magnitude, the strength of the transit remains almost constant. Starting at around 10th magnitude, the SNR begins to drop until decaying to zero near 15th magnitude as the noise components dominate.  Another example in the right panel of Figure \ref{Mag.fig} $-$ KIC 010019708 with a short period and somewhat weaker depth $-$ shows similar results.

The blue vertical lines in Figure \ref{Mag.fig} give the magnitude where TCF SNR drops below 20, setting a rough threshold for detectability.  These threshold magnitudes for all 19 sample stars are given in the 6th column of Table \ref{Kepler19.tbl}.  Most thresholds are fainter than the original Kepler magnitude, indicating the planets could be recovered from the ground from these specific Kepler stars by a ground-based survey with HATSouth noise and irregular cadence characteristics.

These results validate that the ARPS methods recovers planet signals for stars observed with ground-based cadences and noise levels for 4 consecutive seasons for transits represented by Kepler stars in Table~\ref{Kepler19.tbl}; that is, providing the transit depths $\gtrsim 0.1$\%, periods $\lesssim 40$ days, and stellar magnitudes $\lesssim 15$.  Some failures are seen for fainter stars with weaker transits; these effects are investigated below.

We recall that transit depth for a given planet increases as stellar radius decreases, and the distribution of stellar radii among target stars has a magnitude dependence, with many more smaller radius M dwarfs in the sample at magnitudes fainter than 14th.
This effect is not treated here, and could result in more transit detections at fainter magnitudes than presented here.

\subsection{Light Curve Cadence and Duration Dependence}\label{Detect_dur.sec}

An important parameter of ground-based surveys is whether the structure of observations is confined to a single site at non-polar latitudes $-$ roughly 8 hrs per day $-$ or are capable of longer continuous observations $-$ 16 or 24 hours per day $-$ from telescope sites at multiple longitudes or polar location.  Longer observations per day should result in a stronger value of TCF SNR and therefore a more detectable planet, give a fixed transit period and depth. To quantify this effect, tests are run by replacing Kepler data points with NA values to emulate recording data for 8 hr, 16 hr, and 24 hr daily cadence.

\begin{figure}
\centering
\includegraphics[width=1.0\textwidth]{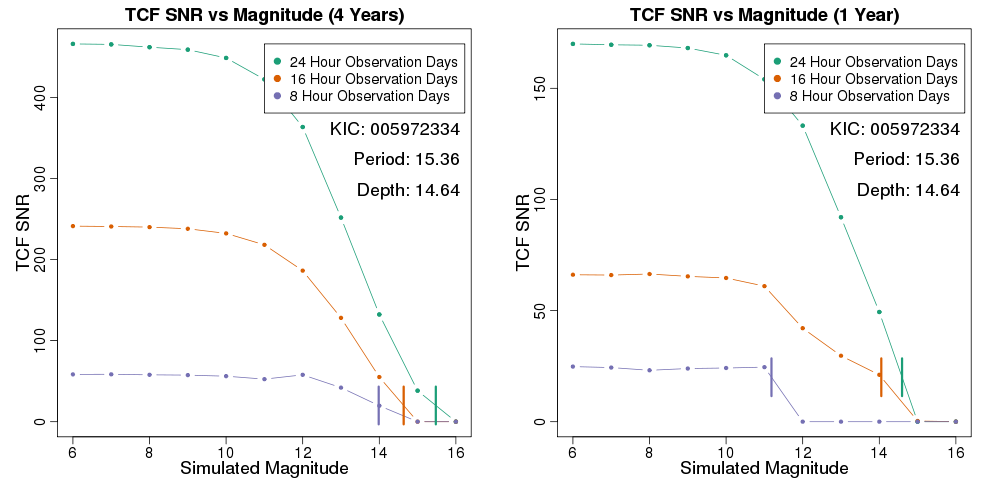}
\caption{The left panel plots TCF SNR for different amounts of observations per day versus magnitude on 4 years of simulated data for KIC 005972334. The right panel depicts the same information for 1 year of data. The thick vertical lines indicate an estimate of when SNR drops below 20. \label{005972334_Cad.fig}}
\end{figure}

\begin{figure}
\centering
\includegraphics[width=1.0\textwidth]{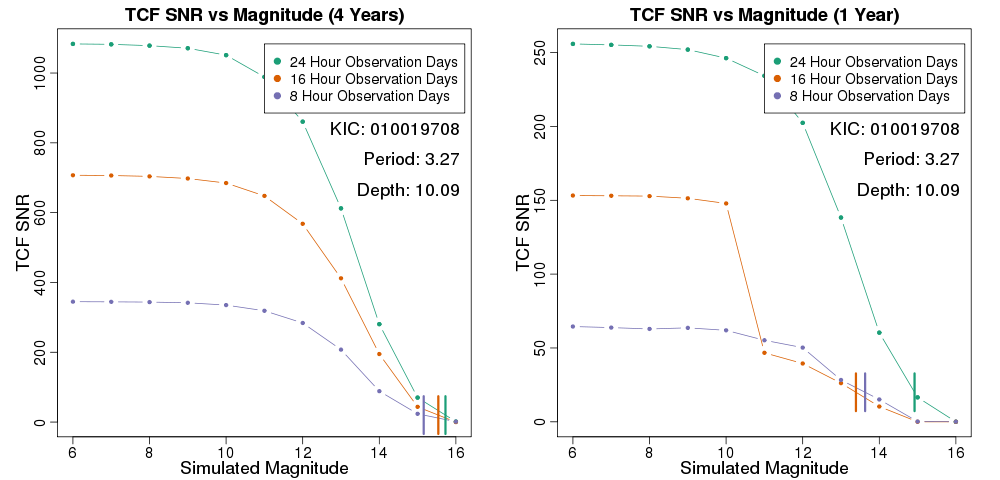}
\caption{The left panel plots TCF SNR for different amounts of observations per day versus magnitude on 4 years of simulated data for KIC 010019708. The panels are similar to Figure~\ref{005972334_Cad.fig}. \label{010019708_Cad.fig}}
\end{figure}

Figure \ref{005972334_Cad.fig} shows the TCF periodogram SNR results for the long-period KIC 005972334 versus simulated magnitude for the three cadence patterns and two light curve durations (4 years and 1 year).  The plots show a sharp degradation in SNR when observation duration is only 1 years and cadences are only 8 hours d$^{-1}$.    Signals are greatly enhanced with continous 24 hr cadences.  This particular planet signal can be detected for stars as faint as $14-16$ mag for all three cadences assuming 4 years observation, but detection is limited to stars brighter than $\sim 11$ mag for the 8-hr cadence and only 1 year (actually 6 months) observation. 

Similar findings are shown in Figure \ref{010019708_Cad.fig} for the short-period planet KIC 010019708.  Planet detectability decays with fewer observations, though the decay is slower than in Figure \ref{005972334_Cad.fig}. When transitioning to one year of data, having twenty four hour observation days becomes more important with 8 hour observation days and 16 hour observation days becoming almost indistinguishable at fainter magnitudes. A continuous 24 hour cadence results in an SNR about five times stronger than when using fewer observations per day. 

For the full samples of 19 deep-transit Kepler stars, the trends for cadence pattern simulations are similar.  Details are reported in Table~\ref{Kepler.tbl} giving S/N ratio of the TCF periodogram at the expected planet transit period for 19 Kepler stars simulated with a range of magnitudes, cadence densities, and survey durations. With few exceptions, the more observations in the sample, the greater the detectability of a star's planetary transit. Unsurprisingly, ARPS modeling works best on data sets with more observations per day and for surveys of longer duration. 

\begin{deluxetable}{rrrrrrrrr}
\tablecaption{TCF periodogram S/N ratios for simulated ground-based observations of 19 Kepler planets \label{Kepler.tbl}}  
\tabletypesize{\footnotesize}
\tablewidth{0pt}
\centering
\tablehead{
\colhead{KIC} & \multicolumn{8}{c}{Simulated properties} \\ \cline{2-9}
 && \multicolumn{3}{c}{Cadence (1 y)} && \multicolumn{3}{c}{Cadence (4 y)} \\ \cline{3-5} \cline{7-9}
 & & 8 h & 16 h & 24 h && 8 h & 16 h & 24 h  \\ 
\colhead{(1)} & \colhead{(2)} & \colhead{(3)} & \colhead{(4)} & \colhead{(5)} && \colhead{(6)} & \colhead{(7)} &\colhead{(8)}
}
\startdata
005972334 && {\bf 8} & {\bf 16} & {\bf 22} && {\bf 35} & {\bf 65} & {\bf 89} \\ \hline
& 6 & 25 & 66 & 170 && 58 & 241 & 466 \\
& 7 & 24 & 66 & 170 && 58 & 241 & 465 \\
& 8 & 23 & 66 & 169 && 58 & 240 & 462 \\
& 9 & 24 & 65 & 168 && 57 & 238 & 459 \\
& 10 & 24 & 65 & 165 && 56 & 232 & 449 \\
& 11 & 25 & 61 & 154 && 52 & 218 & 422 \\
& 12 & \nodata & 42 & 133 && 58 & 186 & 363 \\
& 13 & \nodata & 30 & 92 && 42 & 128 & 252 \\
& 14 & \nodata & 21 & 49 && \nodata & 55 & 132 \\
& 15 & \nodata & \nodata & \nodata && \nodata & \nodata & 38 \\
& 16 & \nodata & \nodata & \nodata && \nodata & \nodata & \nodata \\ \hline
\enddata
\tablecomments{This table is published in its entirety for 19 stars in the electronic version of the paper. A portion is shown here for one star to give guidance regarding its form and content. The top line gives the number of transits in bold face, and the remaining lines give the TCF periodogram SNR at the published period.  Entries below the threshold SNR=20 are shown as {\bf ...}. }
\end{deluxetable}

\subsection{Transit Period and Depth Dependence}

With the Kepler stars in Table \ref{Kepler19.tbl} having differing transit depths, one would expect to have higher detectability with ARPS methodology for deeper transits. Similarly, one would expect larger detectability for shorter periods since more transits could be captured in a survey of limited duration. Using the simulations described above, the effect transit depth and period have on SNR is compared. Table ~\ref{Kepler.tbl} gives full details; we restrict comments here to the optimistic case of 6th magnitude stars observed for four years with a continuous 24 hours a day cadence.

Figure \ref{Period_and_Depth.fig} shows a plot of transit depth versus period for the 19 Kepler star sample with data point symbols scaled to represent TCF SNR.  Planet detectability is sensitive to transit depth in the sense that depths $<0.1$\% are missed or only marginally detected. For larger planets, shorter periods are advantageous over longer periods.  Although scatter is present, the trends show that ARPS is most effective at detecting transits with shorter periods and deeper transit depths.

\begin{figure}
\centering
\includegraphics[width=0.5\textwidth]{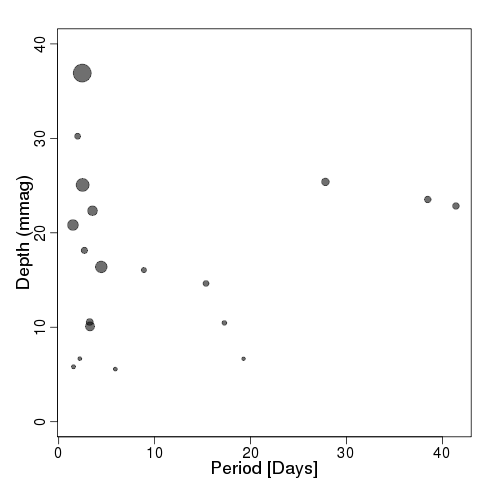}
\caption{Transit depth versus period for the 19 Kepler stars from Table \ref{Kepler19.tbl} with symbol sizes scaled to represent relative TCF SNR for a simulation at 6th magnitude observing for 24 hours a day. The symbol size is thus a measure of recoverability of the planet by the ARPS procedure.} \label{Period_and_Depth.fig}
\end{figure}

\section{Detecting Kepler Stars with Different Cadence Patterns} \label{Patterns.sec}

Section~\ref{Detectability.sec} examined the 19 deep transit Kepler planets detectable from ground-based telescopes, adopting simplistic cadence patterns: evenly spaced observations without gaps within the trial surveys with different daily and annual coverages. Here we examine the effects of realistic unevenly spaced cadences within a day.  For example, we want to learn whether there is a serious loss of information by binning irregular observations into a fixed time sequence. The analysis continues to use the deep Kepler transits, but also applies ARPS methods to planetary signals discovered in the HATSouth survey. When Kepler planets are used, the simulations assume a simulated star brightness of 6th magnitude so that the ground-based noise levels is negligible in order to isolate discovery dependence on cadence pattern alone.

\subsection{Kepler Stars with HATSouth Cadences} \label{HATSCad.sec}

The observing times of the 34 HATSouth transits with confirmed planets (Table~\ref{HATS34.tbl}) are binned into 29.4 minute time intervals using the method detailed in \S\ref{Method_irreg.sec} in order to mimic the Kepler data set. Then an artificial light curve is created for each Kepler star using Kepler fluxes in time slots covered by the HATSouth irregular cadence patterns. Time slots missing from the HATSouth cadence are designated to have `NA' fluxes. Noise components are added using equation \ref{noise} simulated at 6th magnitude. This procedure gives Kepler data with cadence patterns similar to a variety of HATSouth stars.  Steps in the construction of this synthetic light curve is illustrated in Figure \ref{HATS_Cadence_Example.fig}.  

We then run the ARIMA fitting and TCF periodogram stages of the ARPS methodology on the resulting light curves for the 19 stars in Table \ref{Kepler19.tbl} over the 34 different cadences provided by the HATSouth stars. Results are summarized in two tables:  Table \ref{HATS_Cadence.tbl} gives the success rates for different Kepler stars for the 34 cadences, and Table~\ref{HATS_Planets.tbl} gives the success rate of different HATSouth cadences for the 19 stars. The results are listed in order of most successful to least successful planet recoveries.  

The findings here extend those found in in \S\ref{Detectability.sec}. Table \ref{HATS_Cadence.tbl} shows that short periods and deep depths provide the best recovery rate for different HATSouth cadences. The ARPS methodology detects planets for more than half of the cadences when the periods are $\lesssim 4$ days and transit depths $\gtrsim1.5$\%.  Table~\ref{HATS_Planets.tbl} shows that every HATSouth irregular cadence had recoverable planets for some Kepler stars, but no cadence detects more than $\sim 2/3$ of the Kepler planets.  The more successful cadences have missing data in $65-85$\% of the time slots, but planet detections tend to fail when cadences have  $>90$\% missing data. 

Figure \ref{HATS_Cadence_Example.fig} illustrates this range of success for one Kepler star showing the cadence of HATS-24 (with 95\% missing data and 1121 time slots filled) and the cadence of HATS-30 (with 84\% missing data and 4394 time slots filled). The dramatic difference in recovery between these two cadences (14\% $vs.$ 45\% of Kepler planets) is due to the richness of data points. HATS-30 has only 47\% of the data are represented by NA's in its most dense region (around $400-500$ days). In contrast, the cadence pattern for HATS-24 has 90\% missing data in its densest data region (around $300-400$ days).   

Combining Tables~\ref{HATS_Cadence.tbl}-\ref{HATS_Planets.tbl}, we find that planet recovery under the ARPS methodology works best for planets with periods $\lesssim 4$ days and transit depths $\gtrsim 1.5$\% that are observed with rich cadence patterns with $\gtrsim 3000$ time slots filled and missing data fractions under 85\%.

\begin{deluxetable}{lrrrclrrr}
	\tablecaption{Success Rate of Detecting Kepler Planets Using HATSouth Cadences \label{HATS_Cadence.tbl}}
	\tabletypesize{\footnotesize}
	\tablewidth{0pt}
	\centering
	\tablehead{
		\colhead{KIC} & \colhead{Period} & \colhead{Depth} & \colhead{Success} & \colhead{~~}& 
		\colhead{KIC} & \colhead{Period} & \colhead{Depth} & \colhead{Success}  \\
		& \colhead{day} & \colhead{mmag} & \colhead{\%} && & \colhead{day} & \colhead{mmag} & \colhead{\%}
	}
	\startdata
	005794240 & 2.455 & 36.9~ & 94~~~~ && 	010666592 & 2.205 & 6.67 & 24~~~~ \\
	010619192 & 1.486 & 20.8~ & 94~~~~ && 	011449844 & 38.479 & 23.5~ & 24~~~~ \\
	009651668 & 2.684 & 18.1~ & 85~~~~ && 	005972334 & 15.359 & 14.6~ & 12~~~~ \\
	003749365 & 1.974 & 30.2~ & 82~~~~ && 	006522242 & 41.408 & 22.8~ & 12~~~~ \\
	005358624 & 3.526 & 22.3~ & 79~~~~ && 	000757450 & 8.885 & 16.1~ & 6~~~~ \\
	010019708 & 3.269 & 10.1~ & 68~~~~ && 	010790387 & 117.931 & 8.51 & 0~~~~ \\
	011804465 & 4.438 & 16.4~ & 53~~~~ && 	011187436 & 5.907 & 5.57 & 0~~~~ \\
	011517719 & 2.496 & 25.1~ & 50~~~~ && 	009946525 & 51.847 & 9.64 & 0~~~~ \\
	006061119 & 27.808 & 25.4~ & 41~~~~ && 	002987027 & 17.276 & 10.5~ & 0~~~~ \\
	004570949 & 1.545 & 5.82 & 35~~~~ && 		003323887 & 19.271 & 6.66 & 0~~~~ \\
	010874614 & 3.235 & 10.6~ & 32~~~~ && 	003832474 & 143.206 & 10.6~ & 0~~~~ \\
	\enddata
\end{deluxetable}

\begin{deluxetable}{lrrrclrrr}
	\tablecaption{Success Rate of HATSouth Cadences for Detecting Kepler Planets \label{HATS_Planets.tbl}}
	\tabletypesize{\footnotesize}
	\tablewidth{0pt}
	\centering
	\tablehead{
		\colhead{KIC} & \colhead{N Points} & \colhead{NA} & \colhead{Success} &\colhead{~~}&
			\colhead{KIC} & \colhead{N Points} & \colhead{NA} & \colhead{Success}\\
		& & \colhead{frac} & \colhead{\%} && & & \colhead{frac} & \colhead{\%}
	}
	\startdata
	HATS-4 & 4532~~ & 0.84 & 64~~~~ && 	HATS-26 & 1660~~ & 0.72 & 32~~~~ \\
	HATS-28 & 4270~~ & 0.65 & 59~~~~ && 	HATS-15 & 2491~~ & 0.86 & 32~~~~ \\
	HATS-23 & 4281~~ & 0.65 & 59~~~~ && 	HATS-8 & 2186~~ & 0.78 & 32~~~~ \\
	HATS-33 & 4300~~ & 0.85 & 50~~~~ && 	HATS-2 & 2312~~ & 0.76 & 32~~~~ \\
	HATS-31 & 3123~~ & 0.71 & 50~~~~ && 	HATS-27 & 2351~~ & 0.90 & 27~~~~ \\
	HATS-29 & 3870~~ & 0.87 & 50~~~~ && 	HATS-21 & 2876~~ & 0.90 & 27~~~~ \\
	HATS-14 & 2721~~ & 0.85 & 50~~~~ && 	HATS-20 & 2372~~ & 0.90 & 27~~~~ \\
	HATS-3 & 2883 ~~& 0.84 & 50~~~~  && 	HATS-17 & 2329~~ & 0.90 & 27~~~~ \\
	HATS-30 & 4394~~ & 0.84 & 45~~~~ && 	HATS-10 & 2155~~ & 0.94 & 27~~~~ \\
	HATS-19 & 3309~~ & 0.87 & 45~~~~ && 	HATS-5 & 1646~~ & 0.92 & 27~~~~ \\
	HATS-13 & 2362~~ & 0.87 & 45~~~~ && 	HATS-12 & 1952~~ & 0.94 & 23~~~~ \\
	HATS-22 & 2976~~ & 0.93 & 41~~~~ && 	HATS-11 & 1981~~ & 0.94 & 23~~~~ \\
	HATS-18 & 2972~~ & 0.93 & 41~~~~ && 	HATS-9 & 1952~~ & 0.94 & 23~~~~ \\
	HATS-6 & 2403~~ & 0.92 & 41~~~~  && 	HATS-7 & 1610~~ & 0.86 & 23~~~~ \\
	HATS-34 & 2211~~ & 0.67 & 36~~~~ && 	HATS-35 & 2180~~ & 0.92 & 14~~~~ \\
	HATS-32 & 1949~~ & 0.92 & 36~~~~ && 	HATS-25 & 1339~~ & 0.82 & 14~~~~ \\
	HATS-16 & 2669~~ & 0.69 & 36~~~~ && 	HATS-24 & 1121~~ & 0.95 & 14~~~~ \\
	\enddata
\end{deluxetable}

\begin{figure}
	\centering
	\includegraphics[width=0.85\textwidth]{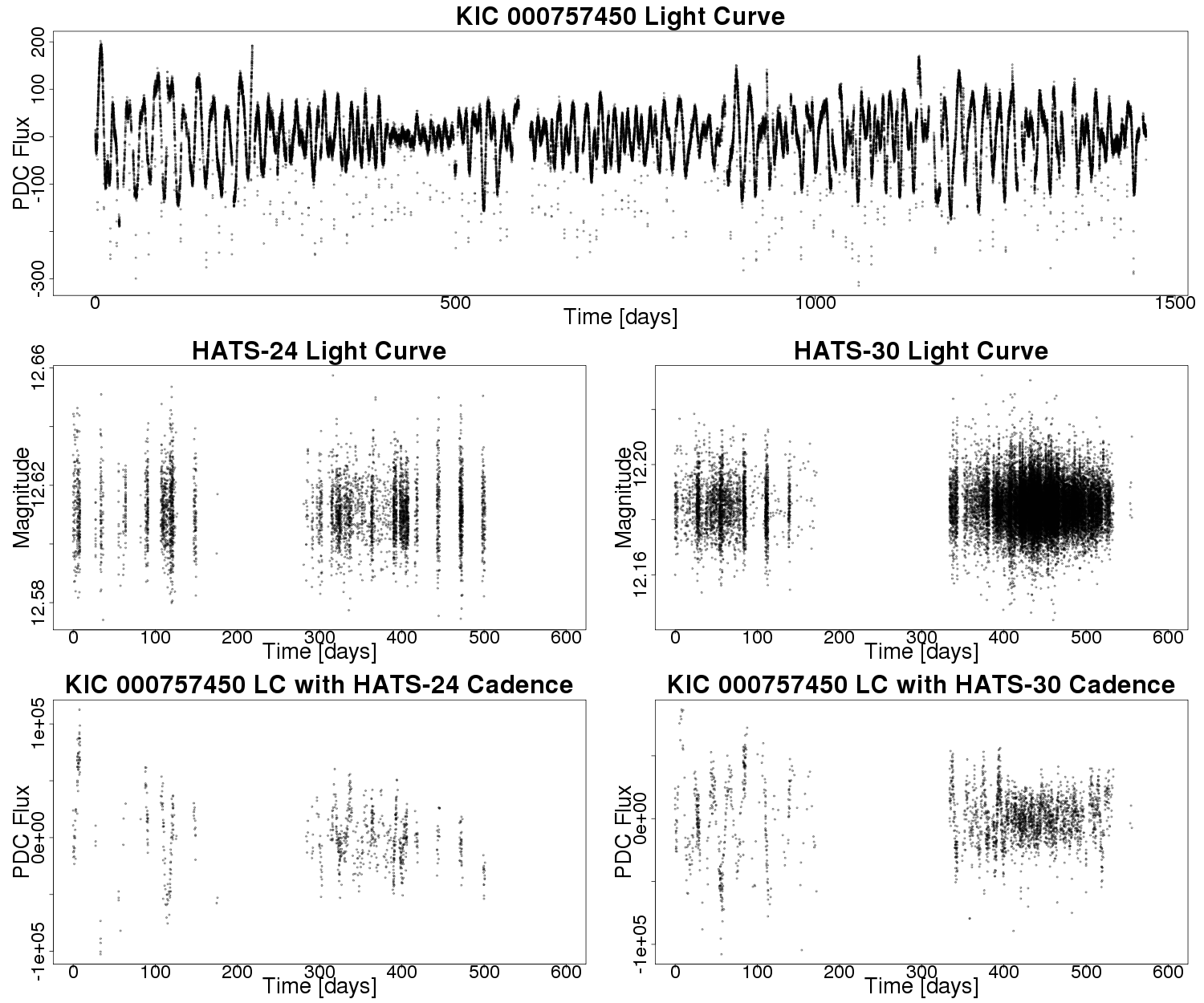}
	\caption{Simulation of a planet-bearing Kepler lightcurve with HATSouth observing cadences. The top panel depicts the light curve for the Kepler Star with KIC 000757450 with strong planetary transits at P = 8.885 days superposed on the stellar variations due to rotationally modulated starspots. The left-center panel shows the original light curve for HATS-24, and the left-bottom panel shows a light curve with values from KIC 000757450 with a sparse cadence pattern corresponding to HATS-24 and flux units amplified to simulate a brighter magnitude used for ARPS analysis. The right-center and right-bottom panels use the dense cadence pattern of HATS-30.  The TCF periodogram did not detect the planet with the HATS-24 cadence but did detect it with the HATS-30 cadence. \label{HATS_Cadence_Example.fig}}
\end{figure}

\section{Detectabilitiy of HATSouth Planets with HATSouth Cadences}  \label{CAD.sec}

Here we leave the Kepler deep-transit sample (Table \ref{Kepler19.tbl}) and examine the effectiveness of ARPS methodology for recovering confirmed planets in the HATSouth survey (Table \ref{HATS34.tbl}).  HATSouth photometry and cadences are given online\footnote{\url{https://hatsouth.org/planets/lightcurves.html}}.  We treat the photometry after instrumental and atmospheric corrections are applied, such as the External Parameter Decorrelation and Trend Filtering Algorithm. This analysis may not represent an unbiased view of the HATSouth planetary population, as it is limited to planetary transits found using the HATSouth data processing system \citep{Bakos13, Bakos18}.

We first bin the HATSouth data into 29.4 minute time intervals to mimic the Kepler data; the ARIMA and TCF stages of ARPS methodology are then applied. Figure \ref{HATS.fig} shows an example for HATSouth star HATS-16a at various stages in the ARPS process. We see in the autocorrelation function (ACF) that the original photometry exhibits strong autoregressive properties. The ACF for the ARIMA residuals, the autocorrelation is much reduced and the noise (measured with the Interquartile Range or IQR) decreases by a factor $>$2.  This demonstrates considerable  effectiveness of the ARIMA model for removing atmospheric and instrumental variations in this star. The TCF recaptures the planetary signal at the correct period of 2.68 days, although spectral peaks of comparable intensity appear at longer periods. It thus would be difficult to identify the planetary signal from the TCF periodograms alone, in this case. The folded light curve shows the box-like transit and the folded ARIMA residuals show the double spike captured by TCF around phase 0.4.

Table \ref{HATS_Recovery.tbl} shows the peak TCF signal-to-noise ratio for each HATSouth planet. ARPS methodology `detected' the known signal for 26 of 34 (75\%) of the published HATSouth planets in the sense that the known period has the highest SNR in the TCF periodogram. However, in many cases, spectral peaks of comparable intensity are present; only half of the planets had a TCF SNR greater than 9. We do not see strong links between ARPS recovery and planet period, transit depth and missing data.  For example, the fraction of missing data is not a good predictor of TCF SNR; some of the successful recoveries have $>90$\% NA fraction and others $<60$\% fraction.  {From the mathematics of ARIMA modeling (equation~\ref{ARMA}), we can infer that ARPS sensitivity will deteriorates with many short gaps but should not be badly affected by long gaps. ARIMA removes unwanted structure on short timescales that will be affected by short gaps, while a 6 month annual gap simple reduces the number of data points under study without affecting the regression fit.

Note that the full AutoRegressive Planet Search procedure does not rely on just the TCF periodogram peak for transit detection, but compares the combined effect of several dozen `features' (from the light curve, ARIMA fit, TCFperiodogram, folded light curve, and stellar metadata) for planet and non-planet training sets using a machine learning Random Forest classifier (Paper I). The ARPS planet recovery rate is likely to be substantially better than those found here using the TCF peak alone. 

\begin{deluxetable}{lrrrrclrrrr}
\tablecaption{HATSouth planets and their corresponding TCF SNR \label{HATS_Recovery.tbl}}
\tabletypesize{\footnotesize}
\tablewidth{0pt}
\centering
\tablehead{
\colhead{Name} & \colhead{Period} & \colhead{Depth} & \colhead{NA} & \colhead{SNR} & \colhead{~~} &
\colhead{Name} & \colhead{Period} & \colhead{Depth} & \colhead{NA} & \colhead{SNR}\\
 & \colhead{days} & \colhead{mmag} & \colhead{frac\tablenotemark{a}} &  &&
 & \colhead{days} & \colhead{mmag} & \colhead{frac\tablenotemark{a}}
}
\startdata
HATS-2 b & 1.35410 & 19.3 & 0.76~~ & 13.0 && HATS-19 b & 4.56967 & 10.3 & 0.87~~ & 19.7 \\
HATS-3 b & 3.57485 & 11.1 & 0.84~~ & \nodata && HATS-20 b & 3.79930 & 8.8 & 0.90~~ & 1.1 \\
HATS-4 b & 2.51673 & 13.9 & 0.84~~ & 9.4 && HATS-21 b & 3.55440 & 13.9 & 0.90~~ & 2.5 \\
HATS-5 b & 4.76339 & 12.6 & 0.92~~ & 0.9 && HATS-22 b & 4.72281 & 22.1 & 0.93~~ & 25.7 \\
HATS-6 b & 3.32527 & 35.1 & 0.92~~ & 13.3 && HATS-23 b & 2.16052 & 27.4 & 0.65~~ & 6.7 \\
HATS-7 b & 3.18532 & 5.5 & 0.86~~ & 1.5 && HATS-24 b & 1.34849 & 18.5 & 0.95~~ & \nodata \\
HATS-8 b & 3.58389 & 7.2 & 0.78~~ & 1.5 && HATS-25 b & 4.29864 & 14.8 & 0.82~~ & 9.4 \\
HATS-9 b & 1.91531 & 5.7 & 0.94~~ & 23.7 && HATS-26 b & 3.30238 & 8.4 & 0.72~~ & 1.7 \\
HATS-10 b & 3.31285 & 8.9 & 0.94~~ & 19.9 && HATS-27 b & 4.63704 & 8.7 & 0.90~~ & 1.2 \\
HATS-11 b & 3.61916 & 12.6 & 0.94~~ &  \nodata && HATS-28 b & 3.18108 & 19.2 & 0.65~~ & 32.8 \\
HATS-12 b & 3.14283 & 4.3 & 0.94~~ & 0.9 && HATS-29 b & 4.60587 & 15.6 & 0.87~~ & 28.4 \\
HATS-13 b & 3.04405 & 21.3 & 0.87~~ & 48.1 && HATS-30 b & 3.17435 & 14.0 & 0.84~~ & 60.3 \\
HATS-14 b & 2.76676 & 14.2 & 0.85~~ & 21.3 && HATS-31 b & 2.54955 & 9.0 & 0.71~~ & 0.9 \\
HATS-15 b & 1.74749 & 16.4 & 0.86~~ & 30.3 && HATS-32 b & 2.81265 & 14.9 & 0.92~~ & 0.2 \\
HATS-16 b & 2.68650 & 12.5 & 0.69~~  &19.9 && HATS-33 b & 2.54955 & 16.6 & 0.85~~ & 70.1 \\
HATS-17 b & 16.25461 & 5.7 & 0.90~~ & \nodata && HATS-34 b & 2.10616 & 24.4 & 0.67~~ & 13.0 \\
HATS-18 b & 0.83784 & 19.7 & 0.93~~ & 74.5 && HATS-35 b & 1.82099 & 12.0 & 0.92~~ & 5.5 \\
\enddata
\tablenotetext{a}{The symbol $...$ represents a TCF peak with power smaller than its local median}
\end{deluxetable}

\begin{figure}
	\centering
	\includegraphics[width=1\textwidth]{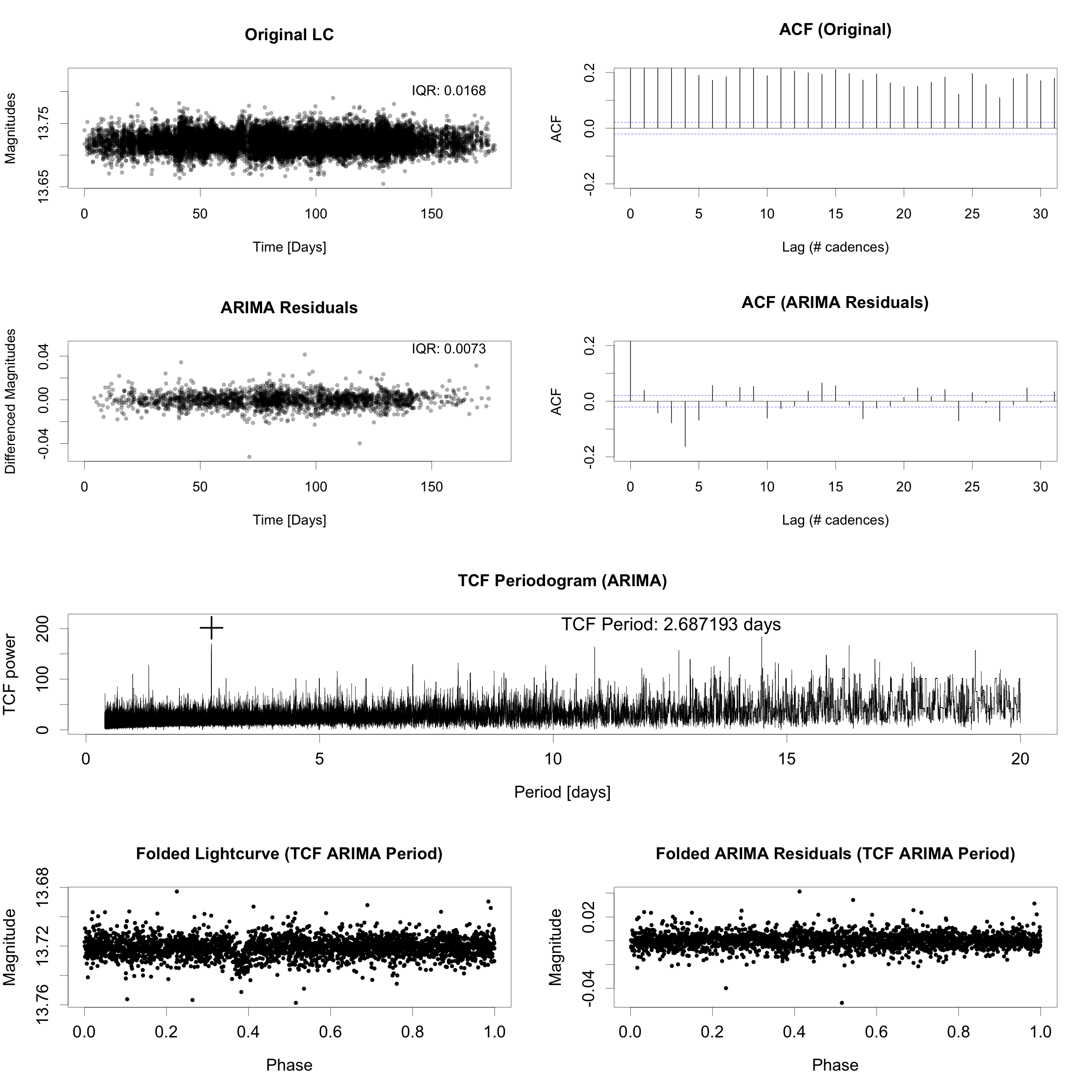}
	\caption{Recovery of HATSouth planet HATS-16b with ARPS methodology. The top row shows the original HATSouth light curve for HATS-16 and its corresponding ACF. The second row shows the ARIMA residuals for the light curve and its corresponding ACF.  The third row shows the TCF periodogram for the ARIMA residuals with the published period marked. The bottom row shows show the folded light curves for the original data and the ARIMA residuals. The transit appears as a box in the left panel and as a double spike in the right panel.  \label{HATS.fig}}
\end{figure}

\section{Discussion}

The central result is that the ARPS statistical analysis procedure for uncovering planetary transits from photometric light curves can be effectively applied to ground-based data with irregular observing cadences.   This was not obvious when we initiated the effort. Parametric autoregressive modeling is specifically designed for evenly spaced time series\footnote{
Statistical modeling of completely irregular time series with autocorrelation is a specialized branch of time series analysis; only one volume is devoted to the topic \citep{Parzen83}.  Parametric methods for such cases include state space modeling \citep{Shumway17} and continuous ARMA (CARMA) processes \citep{Brockwell01, Kelly14}.  Note however that CARMA assumes stationarity (no trends) in the data, and the more general CARFIMA model is mathematically and computationally complex \citep{Tsai05, Tak17}.   A brief review of continuous-time autoregressive modeling of astronomical irregular time series is given by  \citet{Feigelson18}; see also the astronomical discussion by\citet{Hanif15}.}  
Rather we take advantage that maximum likelihood ARIMA  fitting procedures accept `missing data', and we convert the irregular cadences typical of ground-based surveys to evenly spaced cadences with missing data.

We learn that the fraction of cadence times without photometric measurements can be high, with missing data in 50\% to $>$ 90\% of the cadence slots. We do not know any prior studies of ARIMA modeling for evenly spaced time series with such heavy concentrations of missing data.}  It seemed quite possible that the autoregressive model would fail to capture the true character of the variations of the star or observational conditions when the missing data fraction is high.  If the observations are so sparse that the system has undergone unpredictable variations between observations, then any statistical modeling procedure (parametric or nonparametric, low- or high-dimensional) will have dubious scientific value. 

It is therefore satisfying that, in most cases,  the ARIMA models significantly reduce the structure in irregularly spaced stellar lightcurves examined here with a few parameters, and that the TCF periodogram captures the known planet transit signal in the ARIMA residuals (Figures~\ref{example_sim.fig} and \ref{HATS.fig}).  ARIMA modeling and the TCF periodogram are basically successful in detecting planetary signals under these circumstances.  Our investigation then focuses on details regarding how planet detectability decays as cadence pattern becomes sparser and observational noise levels rise.  Our study of light curves with strong planetary transits from the 4-year Kepler mission shows that ARPS methodology is most effective for planets with periods $\lesssim 4$ days and transit depths $\gtrsim 1.5$\% that are observed with rich cadence patterns with $\gtrsim 3000$ time slots filled and missing data fractions under 85\% (\S\ref{Detectability.sec}-\ref{Patterns.sec}).   ARPS recovers three-quarters of HATSouth confirmed planets in the sense that the peak signal of the TCF periodogram has the same period as the published planetary orbit (\S\ref{CAD.sec}). 

We emphasize that the low-dimensional parametric ARIMA approach to noise reduction is mathematically completely different from nonparametric approaches, and the TCF periodogram is comparably or (for short periods) more sensitive than the BLS periodogram (Papers I and II).  Thus when only a tiny fraction of lightcurves from ground-based surveys are shown to have planetary transits by one method, it is quite possible that mathematically different methods will uncover a tiny fraction of new planets with different time series characteristics.   For example, in cases where the original light curve has strong signals in the autocorrelation function at short lags (such as Figure~\ref{HATS.fig}, upper right panel), we expect ARPS to perform better than a method that does not seek to reduce non-planetary signals. Indeed, Paper II shows that the ARPS procedure, with the use of a machine learning classifier trained on a large sample of Kepler confirmed planets, can recover 97\% of known planets above a low threshold and report dozens of new candidate planets with similar transit-like characteristics.

We can roughly estimate how many planets might be found if ARPS were applied to the HATSouth survey.  Section~\ref{Detectability.sec} shows that $\sim 10$ out of $\sim 200,000$ Kepler stars could have been detected with HATSouth providing the stellar magnitude is brighter than $\sim 15$.  If there are about 3 million sufficiently bright stars with reasonably dense cadences in the HATSouth survey \citep{Bakos18}, then ARPS is expected to detect about $\sim 150$ planets, similar to the number found to date using traditional analysis methods.  

But this estimate could be low for several reasons.  First, HATSouth has been operating for 9 years, longer than the 4 year duration assumed in our simulations. {Longer light curve durations can substantially improve the detectability of periodic transients; compare the ordinate values on the left and right panels of Figures~\ref{005972334_Cad.fig}-\ref{010019708_Cad.fig}.  Second, HATSouth will have a higher fraction of small-radius M stars where a given planet produces a deeper transit than in the simulations of Kepler stars (\S\ref{Detect_man.sec}).  Third, the detectability indicator here is based on an arbitrary threshold of a single variable, the signal-to-noise ratio of the TCF periodogram peak ($SNR>20$).  A machine learning classifier based on many `features' of the dataset will be considerably more sensitive than using the TCF peak alone, as found in the Kepler dataset (Paper II).  It is thus quite plausible that an ARPS analysis of the HATSouth dataset would emerge with considerably more than 150 candidate planetary transits, even while failing to capture all of the previously known planets.  

\section{Conclusion}

The ARPS methodology based on ARIMA modeling of non-planetary variability and TCF periodograms for uncovering periodic planetary transits (Paper I) is capable of detecting transits in irregularly spaced ground-based light curves. The detectability of transits using ARPS methodology (Paper I) depends critically on various characteristics of a light curve, as detailed in the simulations described in \S\ref{Detectability.sec}-\S\ref{CAD.sec}.  Results include:
\begin{enumerate}

\item With our assumed noise model and typical HATSouth cadences, ARPS processing recovers Kepler planets with $\geq 1$\% depth for stars brighter than 14th magnitude (Figures \ref{010019708_Cad.fig}-\ref{Period_and_Depth.fig}).  Signal recovery is more difficult with shallower transit depths and longer orbital periods.  Sensitivity to different periods can not be summarized in a simple fashion.  For 24-hour cadences and 4 year light curve duration, ARPS readily recovers Kepler planets with periods up to 40 days for depths $\geq 1$\% (Figure \ref{Period_and_Depth.fig}).  Recovery for longer periods is likely, but the Kepler sample has no long-period deep-depth ($>20$ mmag) planets for testing.  

\item Denser cadences and longer light curve durations increase transit detectability.   Observing for 24-hours each day, either using multiple telescopes different longitudes (as with HATSouth) or at the South Pole (as with AST-3), gives a very strong advantage over a single telescope 8-hour daily cadence (Figures \ref{005972334_Cad.fig}-\ref{010019708_Cad.fig}).  A denser daily cadence is particularly important for shorter duration light curves.   

\item Transit signal recovery deteriorates as the observational noise components become stronger than the transit depth (\S\ref{Detectability.sec}). TCF SNR most often falls below a detectability threshold between magnitudes 15 and 16. At this point, noise components from the simulated noise model (equations \ref{noise}) reach a value of ~30 mmag, dominating transit depth.  This particular noise model is characteristic of HATSouth and will differ for other telescope systems. 

\item ARPS sensitivity to daily cadence pattern is not reflected in the simple measure of `NA fraction' in Table 5.  It might seem that an ARIMA-type model would be ineffective for a typical HATSouth light curve where measurements are missing for $80-90$\% of evenly-spaced time slots.   But most of the missing data are collected in 6-month-long annual gaps, while the NA fraction within the 6-month observation window is typically $< 70$\%.  The ARPS methodology deteriorates with many short cadence gaps, but is not badly affected by a few long gaps.  

\end{enumerate}

Overall, we find that ARPS works effectively on ground-based surveys for short period planetary transits, although its sensitivity is not always high. For the Kepler deep-transit sample, the methodology works best on stars with magnitude 15 or greater and with planetary transits of deep depth and/or short period. Periods greater than 50 days are often unrecoverable due to fewer transits recorded. Additional telescopes recording for more than 8 hours a day also creates denser cadences for more significant planet recovery, suggesting around-the-clock coverage through multiple telescopes or a South Pole telescope is very beneficial.

We analyze planet recovery for the HATSouth survey in some detail.  Tables~\ref{HATS_Cadence.tbl}-\ref{HATS_Recovery.tbl} show a broad range of success due to differing cadence densities and varying transit depths and periods. For example, a star with depth of about 20 mmags had a success rate ranging from 14\% to 64\% due to varying periods of 38 days and 1.5 days respectively, showing stronger recovery for short periods.  Simple measures like the total missing data fraction can be misleading in predicting planet recovery due to the low importance of long gaps;  it is more important that the light curve has dense regions of data capturing several transits.  

When applied to the  ensemble of millions of HATSouth light curves, the ARPS methodology is predicted to be comparably or more sensitive than existing methods for transiting planet detection. This promising result motivates ARPS application to the full HATSouth dataset.  

\acknowledgements

{\bf Acknowledgements:} ~~  We appreciate the helpful commentary of an anonymous reviewer. The ARPS project is supported at Penn State by NSF grant AST-1614690 and NASA grant 80NSSC17K0122. J.H. acknowledges support from NASA grant NNX17AB61G.

\end{document}